# Imaging and control of critical spin fluctuations in two-dimensional magnets

Chenhao Jin[1*], Zui Tao[1,2], Kaifei Kang[2], Kenji Watanabe[3], Takashi Taniguchi[3], Kin Fai Mak[1,2,4*], Jie Shan[1,2,4*]

[1] Kavli Institute at Cornell for Nanoscale Science, Ithaca, NY 14853, USA

[2] School of Applied and Engineering Physics, Cornell University, Ithaca, NY 14853, USA

[3] National Institute for Materials Science, 1-1 Namiki, Tsukuba, 305-0044, Japan.

[4] Laboratory of Atomic and Solid State Physics, Cornell University, Ithaca, NY 14853, USA.

* Correspondence to: jinchenhao@cornell.edu, km627@cornell.edu, jie.shan@cornell.edu

**Abstract:**

Strong spin fluctuations are expected near the thermodynamic critical point of a continuous magnetic phase transition. Such critical spin fluctuations are highly correlated and in principle can occur at any time- and length-scales[1]; they govern critical phenomena and potentially can drive new phases[2,3]. Although theoretical studies have been made for decades, direct observation of critical spin fluctuations remains elusive[2,3]. The recent discovery of two-dimensional (2D) layered magnets, in which spin fluctuations significantly modify magnetic properties as compared to their bulk counterparts and integration into heterostructures and devices can be easily achieved, provides an ideal platform to investigate and harness critical spin phenomena. Here we develop a fast and sensitive magneto-optical imaging microscope to achieve wide-field, real-time monitoring of critical spin fluctuations in single-layer $CrBr_3$, which is a 2D ferromagnetic insulator. We track the critical phenomena directly from the fluctuation correlations and observe both slowing-down dynamics and enhanced correlation length. Through real-time feedback control of critical spin fluctuations, we further achieve switching of magnetic states solely by electrostatic gating without applying a magnetic field or a Joule current. The ability to directly image and control critical spin fluctuations in 2D magnets opens up exciting opportunities to explore critical phenomena and develop applications in nanoscale engines and information science.

Two-dimensional (2D) van der Waals magnets have attracted much recent interest[2-19]. As atomically-thin layers, they have been incorporated into van der Waals heterostructures and device architectures to enable novel functionalities, such as gate-controllable magnetism and spin-filter tunneling[7-14]. The 2D nature of layered magnets also leads to distinct physical properties from their bulk counterparts ranging from stacking-dependent magnetism to topological spin excitations[15-19]. Of particular interest are spin fluctuations near a thermodynamic critical point[2,3], which are crucial to understanding magnetism in two dimensions but remain largely unexplored experimentally. In three-dimension the phase space for thermal fluctuations to become critical is small and difficult to access according to the Ginzburg criterion[1,20]. In one-dimension fluctuations are so strong that magnetic long-range order is typically destroyed[21]. 2D layered magnets, in which fluctuations and long-range order reach a good balance, are therefore ideal to access and harness critical spin fluctuations.

Here we demonstrate real-time imaging of critical spin fluctuations and direct determination of their temporal and spatial correlations in 2D ferromagnetic $CrBr_3$ layers by a fast and sensitive magneto-optical imaging technique. We observe macroscopic spatial correlations and orders-of-magnitude change in the spin correlation time within a narrow temperature range near the Curie temperature $T_c$. The observed extreme sensitivity of spin fluctuations to environment is distinctively different from non-critical thermal fluctuations in miniscule systems, and enables unprecedented flexibility for control. We are thereby able to switch the magnetic state of 2D $CrBr_3$ in a non-volatile, magnetic-field-free and current-free manner by harnessing critical fluctuations as stochastic driving forces, i.e. by toggling the critical fluctuations based on a real-time measurement of the magnetic state.

Being stochastic in nature, direct observation of critical fluctuations in real-time for a large sample area is challenging. We have developed a real-time magnetic circular dichroism (MCD) imaging technique that combines high temporal (up to 100 frames per second) and spatial (~ 600 nm) resolution with high sensitivity, which allows us to monitor the spin fluctuations in a single-layer ferromagnet $CrBr_3$. Figure 1a shows an illustration of our experimental setup. It enhances the optical contrast of MCD through polarization control (see supplementary note 2 and methods). The concept of polarization control is well known in polarization microscopy, but the enhancement of the optical contrast is limited by the low polarization extinction ratio, which is typically several hundred due to polarization distortion from high numerical-aperture (NA) objectives[22,23]. The key improvement of our technique is to separate the effective NA for illumination and imaging: A long-focal-length lens L1 focuses the illumination beam roughly at the back aperture of an objective, minimizing the effective NA for illumination and thus polarization distortion. Similarly, specular-reflected light from a featureless substrate also maintains high polarization purity. However, scattered light from features on the sample is collected with the large effective NA of the objective. The spatial resolution is thus not compromised, as there is no physical aperture limiting the NA in the collection path. We are therefore able to achieve simultaneously a high extinction ratio ($> 3\times10^4$) and high spatial resolution (~ 600 nm) over the entire field of view. In this configuration, the optical contrast is approximately proportional to the MCD signal (and therefore also the out-of-plane magnetization of the sample), but is enhanced by more than 100 times compared to the MCD signal without polarization control. In addition, because the MCD signal is determined by the dielectric function of the sample and the local field factor of the environment, we can optimize the substrate to further improve the optical contrast by a few times (see supplementary note 3).

Figure 1b shows an optical microscopy image of a monolayer $CrBr_3$ sample S1 (inside the white dashed box), which is encapsulated with hexagonal boron nitride (hBN) on both sides. It shows a giant magneto-optical contrast of ±60% for the two remnant magnetization states (Fig.1c and 1d), which enables imaging of the magnetization in 10 ms exposure time. A nearby $CrBr_3$ flake (inside the green dashed box) can also be seen, but will not be focused on because of its smaller size. We first characterize the $CrBr_3$ monolayer away from $T_c$. The magnetization against magnetic field shows a hysteresis loop below $T_c$ and a paramagnetic behavior above $T_c$ (Fig. 1e), from which we obtain the coercive field and the DC magnetic susceptibility $\chi$, respectively. The coercive field decreases dramatically when temperature $T$ approaches $T_c$ from below; and $\chi$ increases dramatically when $T$ approaches $T_c$ from above (red circles in Fig. 1f). In contrast, a thin bulk $CrBr_3$ flake (~10 nm thick) shows a much weaker temperature dependence in susceptibility above $T_c$ and no remnant magnetization below $T_c$ (black squares). The lack of remnant magnetization is presumably due to domains or antiferromagnetic ordering from the interlayer exchange and/or dipole-dipole interaction. We fit the temperature-dependent susceptibility with a scaling law $\chi \sim (T-T_c)^{-\gamma}$ in the inset, and obtain $T_c$ = 22.3 K and 28 K, and critical exponent $\gamma$ = 2.4 and 1.2 for the monolayer and thin bulk samples, respectively. The critical exponent $\gamma$ of the thin bulk matches well with previous results for bulk $CrBr_3$, and is close to the mean-field value ($\gamma$ = 1)[24]. On the other hand, the critical exponent for monolayer $CrBr_3$ is between the predications of the 2D Ising ($\gamma$ = 1.75) and 2D Heisenberg ($\gamma$ = 3) model[25,26]. This can potentially be understood from a recent theoretical proposal that 2D $CrBr_3$ is described by a 2D XXZ model with anisotropic exchange interaction[17], and shows a crossover behavior between the Ising and Heisenberg model. Although the exact model to describe 2D $CrBr_3$ is still under debate[17-19], our observation demonstrates its distinct behavior from that of the bulk, and the substantial deviation from the mean-field description due to enhanced fluctuations in 2D.

As temperature further approaches $T_c$ (22.3K) from below, the magnetization in monolayer CrBr$_3$ starts to fluctuate spontaneously under zero external magnetic field, which is not observed in the bulk sample (see supplementary movie 1). Figure 2b shows three representative snapshots of the MCD image at 21.9 K. Regions of positive and negative magnetizations coexist and fluctuate over a large length-scale (see supplementary movie 2-13 for the real-time magnetization fluctuations at different temperatures). Figure 2c represents the magnetization dynamics at a fixed location P1 (green circle in Fig. 2b) for varying temperatures. Particularly, in the vicinity of $T_c$, the magnetization shows random values between that of the fully spin-up state (referred to as state "1") and the fully spin-down state (state "0"). The measured magnetization fluctuations and their temperature dependence match well with Monte Carlo simulation from Ising model[27], confirming their origin from critical spin fluctuations (also see supplementary note 9 for multiple control experiments that exclude experimental artifacts as origins of the observed magnetization fluctuations). Figure 2a summarizes the fluctuation amplitude map of the magnetization, $\delta M(\boldsymbol{r}) = \sqrt{\langle M(\boldsymbol{r},t)^2 \rangle - \langle M(\boldsymbol{r},t) \rangle^2}$, at representative temperatures. Here $M(\boldsymbol{r},t)$ is the out-of-plane magnetization at a given point $(\boldsymbol{r},t)$ in space and time, and the time average $\langle \cdots \rangle$ is equivalent to the ensemble average by assuming ergodicity (see methods). The magnetization fluctuations first emerge at the corners of the monolayer flake at 21.6 K, quickly expand into the center with increasing temperature, remain large until $T_c$, then rapidly diminish above $T_c$.

The central quantity that describes the critical fluctuations is the fluctuation correlation function between magnetizations at point $\boldsymbol{r}_1$ and $\boldsymbol{r}_2$ separated by time $\Delta t$ (Ref.[1])

$$C(\boldsymbol{r_1}, \boldsymbol{r_2}, \Delta t) = \langle M(\boldsymbol{r_1}, t) M(\boldsymbol{r_2}, t + \Delta t) \rangle - \langle M(\boldsymbol{r_1}, t) \rangle \langle M(\boldsymbol{r_2}, t + \Delta t) \rangle. \quad (1)$$

The temporal correlation at a given point, $C(\boldsymbol{r_1} = \boldsymbol{r_2}, \Delta t)$, contains direct information of the dynamics in the critical regime. Figure 2d shows the temporal correlation function at position P1 at varying temperatures. The fluctuations are substantially slowed down at 21.8 K (magenta curve in Fig. 2c), leading to a marked increase in the decay time of the correlation function. Figure 2e summarizes the correlation decay time for position P1 (black symbols) as a function of temperature. The unusually strong critical slowing-down (up to a second) again demonstrates the enhanced critical phenomena in 2D and has been independently confirmed by AC susceptibility measurements (see methods and supplementary note 8 for more discussions). Interestingly, the maximum occurs slightly below $T_c$ (vertical dashed line), not exactly at $T_c$ as in an infinite homogeneous system. Such a deviation is likely caused by the boundary effect. Intuitively, the slowing-down dynamics is related to the increase of the fluctuation length-scale. Boundary, at which the number of neighboring sites is reduced, could facilitate large-scale fluctuations below $T_c$ and explain the observed discrepancy. This picture is supported by Monte Carlo simulation of the 2D Ising model of a finite-size system (see supplementary note 5). It is also consistent with the additional slowing-down around $T_c$ (red symbols in Fig. 2e) observed in the correlation decay time at the center of the flake (P2, cyan circle in Fig. 2b).

The critical behavior of the spin fluctuations and the boundary influences are also clearly seen in the spatial correlation function $C(\boldsymbol{r_1}, \boldsymbol{r_2}, \Delta t = 0)$. Figure 3a shows $C(\boldsymbol{r_1}, \boldsymbol{r_2})$ at several representative temperatures (see supplementary note 4 for more temperatures), where $\boldsymbol{r_1}$ is mapped over the entire field of view and $\boldsymbol{r_2}$ is fixed at position P1 (left panel) or P2 (right panel). Below $T_c$, the correlation functions at both positions are anisotropic with different patterns. The three-fold pattern (P1) and two-fold pattern (P2) at 22.2 K reflect the local geometry of the boundary. Above $T_c$, on the other hand, the patterns become largely isotropic since the boundary effect is less important when the long-range magnetic order is destroyed. Figure 3b and 3c show a line cut of the correlation function map for position P1 and P2, respectively, along the dotted lines in Fig. 3a at varying temperatures. The correlation functions feature both an oscillatory component and an exponential envelope function, and can be well fitted by $\cos(\frac{\pi l}{\lambda}) e^{-\frac{l}{\xi}}$ (solid lines in Fig. 3b and 3c). Here $l$ is the distance to position P1 or P2; and $\lambda$ and $\xi$ represent the oscillation half-period and the decay length, respectively. An enhanced correlation decay length $\xi$ is seen for both positions: near $T_c$ for point P2 at the sample center (red circles in Fig. 3e) and at a slightly lower temperature for point P1 at the flake corner (black squares), presumably again due to the boundary effect (see supplementary note 5 for Monte Carlo simulation results). The oscillatory component in the spatial correlation is not well understood. It suggests the presence of a domain-like structure with a characteristic width $\lambda$ (Fig. 3d). The extracted width $\lambda$ is larger than the sample size for $T$ below ~ 21 K (see supplementary movie 2) and decreases rapidly when $T$ approaches $T_c$. Such a behavior is consistent with the prediction that in 2D ferromagnets even a tiny dipole-dipole interaction becomes increasingly important near $T_c$ and leads to a decreased domain size[28,29].

The sharp temperature dependence of the critical spin fluctuations in 2D magnets demonstrated above is fundamentally different from the behavior of non-critical Brownian fluctuations in miniscule systems[30] and opens up opportunities for their efficient control. We fabricate field-effect devices of monolayer $CrBr_3$ with a graphene gate and contact, and achieve tuning of $T_c$ at a rate of ~ -0.4 K/V by applying a gate voltage $V_g$. Gating primarily introduces doping and tunes the intralayer magnetic exchange coupling in monolayer $CrBr_3$ (See methods and supplementary note 7)[12]. Because critical fluctuations depend sensitively on how close the system is from the critical point, they can be effectively controlled by $V_g$ for instance at a fixed sample temperature. Figure 4a shows the magnetization dynamics of device S2 under different gate voltages at 17.90 K (below $T_c$ at $V_g$=0 V). The corresponding gate-dependent fluctuation amplitude and the correlation decay time are shown in Fig. 4b. At this temperature, application of $V_g$ < 0.5 V can turn on the critical fluctuations that are absent without $V_g$. In Fig. 4c we demonstrate switching of a magnetic bit by harnessing the critical fluctuations. In the unshaded regions ($V_g$=0 V), the critical fluctuations are absent, and the magnetic bit stays either in state "1" or state "0" as prepared. In the yellow shaded regions ($V_g$=0.4 V), the fluctuations are activated. Using real-time feedback control (i.e. removing $V_g$ at the right moment according to the real-time magnetization measurement), one can set the bit into a desired state (see supplementary note 6 for similar switching operation in sample S1 with optical control). Figure 4d further shows that the magnet can be repeatedly and reversibly switched between states "1" and "0" by applying square gate voltage pulses with an amplitude 0.4 V and a pulse width 50 ms (see methods). Because the spontaneous and stochastic fluctuations are the only driving forces here, the outcome cannot be deterministic if without feedback. Deterministic logical operations can be achieved by making one measurement after each gate pulse until the desired state is reached; and in principle the only energy cost of the switching operation is in measuring the system's state. This critical-fluctuation-based concept can therefore potentially provide a solution to efficient magnetic processing and storage, such as logic gates and race-track memory.

Finally we note that complete understanding of the critical fluctuations in 2D $CrBr_3$ will require a sophisticated model that at least accounts for finite magnetic anisotropy, dipolar interaction and the boundary effect in a framework beyond the mean-field theory. It would be interesting to see the potential crossover between different universality classes or the emergence of hidden phases near the critical point[28,29]; and real-time imaging of fluctuations can provide invaluable information to that purpose. Nevertheless, our observation clearly demonstrates an unexplored and unique aspect of 2D layered ferromagnets compared to their bulk counterparts. Combining the opportunity to access the critical fluctuation regime in 2D, the device-compatibility of layered materials to enable electrical readout and manipulation, and the magneto-optical imaging technique to monitor spin fluctuations in real-time, 2D layered magnets provide an attractive platform for studying spin fluctuations and critical phenomena, as well as for fluctuation-based apparatuses such as dissipationless memories, Brownian motors and reservoir computation[31-34].

**Competing interests:**

The authors declare no competing interests.

**Data availability:**

The data that support the findings of this study are available within the paper and its Supplementary Information. Additional data are available from the corresponding authors upon request.

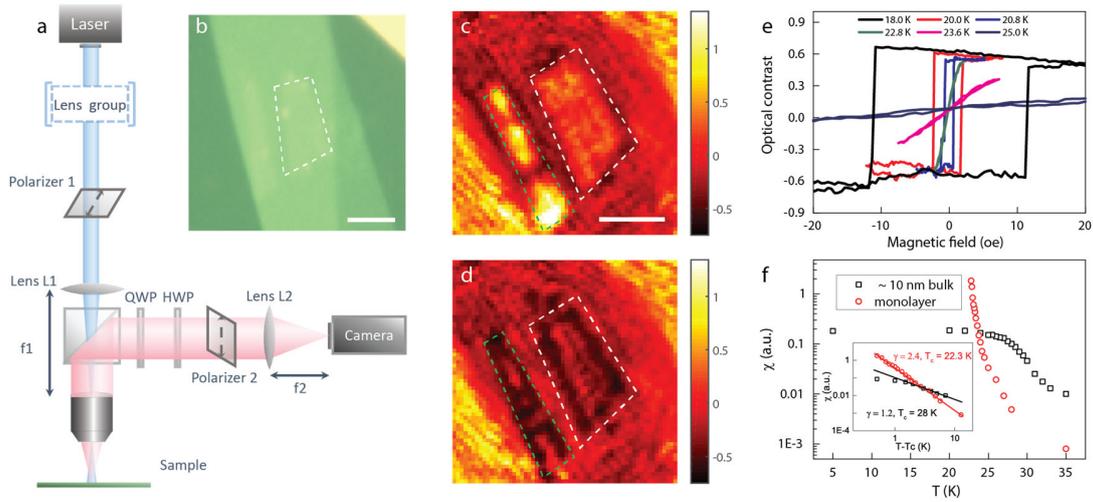

**Figure 1 | Polarization-enhanced MCD imaging of 2D CrBr$_3$. a**, Illustration of the experimental setup. Blue and red beams represent illumination light from the laser and scattered light from the sample. They have different effective NAs. HWP: Half-wave plate. QWP: Quarter-wave plate. **b-d**, Optical microscopy image (**b**) and polarization-enhanced MCD image (**c**, **d**) of a monolayer CrBr$_3$ sample S1 (white dashed box). The MCD image shows giant optical contrast of ±60% for the positive (**c**) and negative (**d**) remnant magnetization. Green dashed box indicates another 2D CrBr$_3$ flake nearby. Scale bar is 5 µm. **e**, Optical contrast of the monolayer against magnetic field shows a hysteresis loop at temperatures below $T_c$ (22.3 K) and paramagnetic behavior above $T_c$. **f**, Temperature dependence of the DC magnetic susceptibility of the monolayer (red) and a bulk of ~10 nm thickness (black), and the corresponding power law fitting (inset). Whereas the bulk shows a behavior close to the mean-field description, the monolayer deviates substantially from it due to the enhanced fluctuations in 2D.

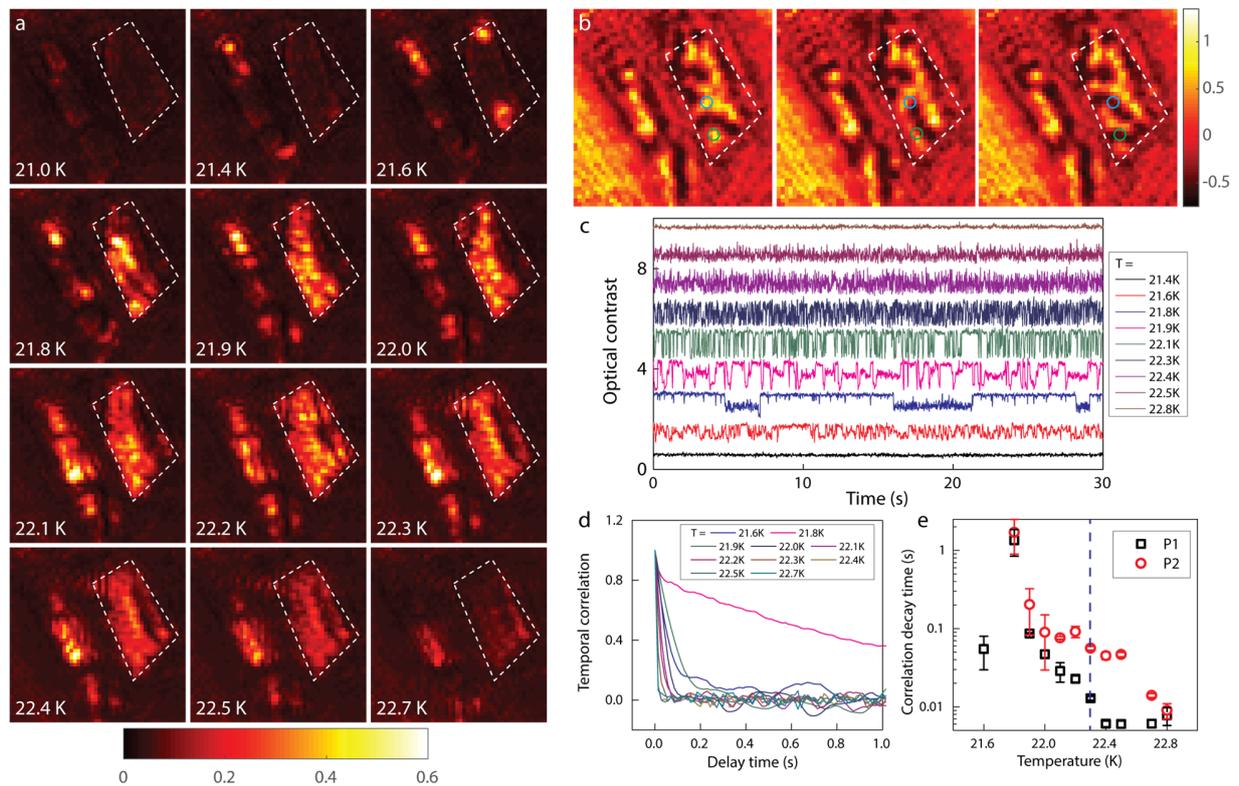

**Figure 2 | Real-time imaging of critical fluctuations in 2D CrBr$_3$. a**, Temperature-dependent amplitude map of critical fluctuations in sample S1. The critical fluctuations emerge first at the corners of the sample and exist in a narrow temperature range of ~ 0.5 K around the critical point (22.3 K). **b**, Snapshots of the critical spin fluctuations at 21.9 K. Patches of positive and negative magnetizations form domain-like patterns that spontaneously change over time. **c**, **d**, Magnetization time trace (**c**) and temporal correlation function (**d**) of position P1 (green circle in **b**) at varying temperatures, showing prominent slowing down around 21.8 K. The magnetization time traces are displaced vertically and successively by 1.25 for clarity. The temporal correlation functions are normalized to 1 at zero delay. **e**, Temperature dependence of the correlation decay time (symbols) and uncertainty (error bars) from exponential function fitting of the temporal correlation function. Compared to P1 at the corner of the sample, the more centered position P2

(cyan circle in **b**) shows additional slowing down near $T_c$ (vertical dashed line), suggesting the importance of the boundary effect.

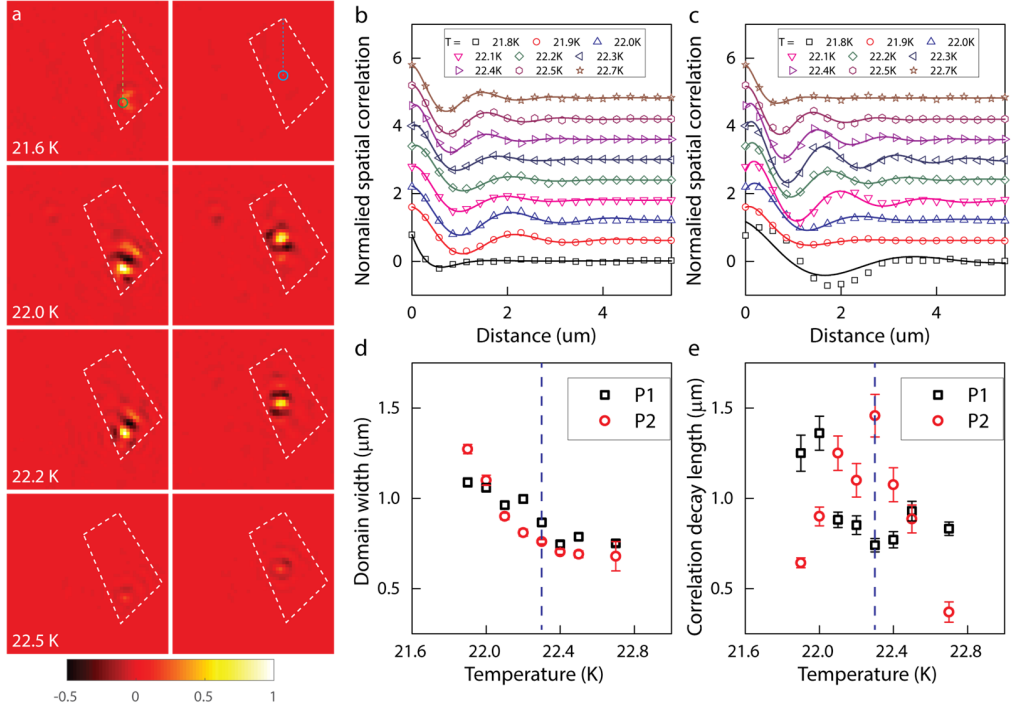

**Figure 3 | Spatial correlation function. a**, Spatial correlation function for position P1 (left) and P2 (right) at representative temperatures. The correlation patterns are anisotropic and distinct between the two points below $T_c$, and largely isotropic and similar above $T_c$. **b**, **c**, Normalized spatial correlation function for P1 (**b**) and P2 (**c**) along the dotted lines in **a**, which are roughly perpendicular to the domain-like structures. Symbols are experiment and solid lines are fits described in the text. Results for different temperatures are displaced vertically and successively by 0.6 for clarity. **d**, **e**, Domain width (**d**) and correlation decay length (**e**) obtained from fitting the spatial correlation function of P1 (black) and P2 (red). Vertical dashed lines represent $T_c$. The characteristic domain width shows a monotonic decrease for both points as temperature approaches $T_c$ from below. The correlation decay length shows an enhancement slightly below $T_c$ for P1 and around $T_c$ for P2, respectively, consistent with the consequence of the boundary effect. Error bars are standard errors of the fitting parameters.

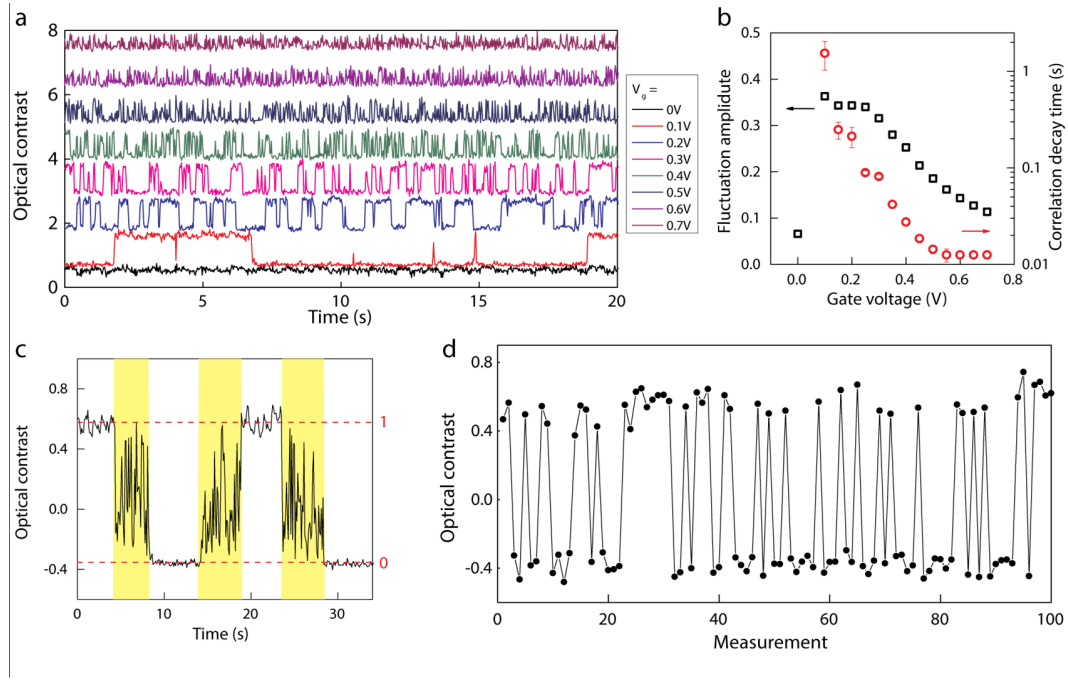

**Figure 4 | Electrical control of the critical fluctuations. a**, Magnetization time traces of position P3 in a monolayer CrBr$_3$ device S2 with graphene back gate. The gate voltage $V_g$ dramatically changes the critical fluctuations through tuning $T_c$. **b**, Gate-dependent amplitude (black) and correlation decay time (red) of critical fluctuations extracted from the time traces in **a**. **c**, At 17.90K (slightly below $T_c$ at $V_g$ =0V), the critical fluctuations are absent without gate voltage (unshaded regions) and the magnetization stays in state "1" or "0" (red dashed lines). A $V_g$ (=0.4V) is used to temporarily turn on the critical fluctuations (yellow shaded regions). A specific state is written into the magnet by removing gate voltage at the right moment according to the real-time magnetization measurements. **d**, Repeated and reversible switching between state "1" and "0" by applying square gate voltage pulses with an amplitude 0.4 V and a pulse width 50 ms at 17.90K (see text and methods for details). A representative sequence of 100 measurements is shown.

**Methods:**

Sample preparation: Thin flakes of CrBr$_3$ encapsulated by hexagonal boron nitride (hBN) on both sides were employed in this study. The hBN-CrBr$_3$-hBN stacks were fabricated using a dry transfer method[35]. CrBr$_3$ and hBN flakes were first exfoliated from their bulk crystals onto silicon substrates with a 300-nm oxide layer. The thickness of hBN and thicker flakes of CrBr$_3$ was determined by atomic force microscopy. The thickness of 2D CrBr$_3$ was determined from the calibrated optical reflection contrast (see supplementary note 1). A stamp made of polycarbonate (PC) on polydimethylsiloxane (PDMS) was used to pick up the top hBN flake, the CrBr$_3$ flake and the bottom hBN flake from Si substrates in sequence at 40°C. For gate-tunable monolayer CrBr$_3$ device S2, two additional bilayer graphene flakes were exfoliated from bulk crystals and picked up during the assembly, serving as the contact and back gate, respectively. The complete structure was then stamped onto a clean silicon substrate with pre-patterned gold electrodes at 120°C to delaminate the PC film with the heterostructure from the PDMS layer. The residual PC was dissolved in chloroform at room temperature. The transfer process was performed inside a nitrogen gas filled glovebox with less than 1 ppm oxygen and moisture to avoid degradation of the CrBr$_3$ flakes.

Real-time MCD imaging: A 409-nm laser diode (Thorlabs L405P150) was used as the probe light source. The photon energy was chosen to be slightly below the absorption edge of the sample to enhance the magneto-optical measurement sensitivity. Two Glan-Taylor polarizers (Thorlabs GT10), one broadband half-wave plate (Thorlabs AHWP05M-600) and one broadband quarter-wave plate (Thorlabs AQWP05M-600) were used to control and analyze the polarization of the probe beam. The probe beam transmitted through a beamsplitter (Thorlabs BS028) and an objective (Olympus LUCPlanFLN 40x, NA = 0.6), and impinged on the sample. The beamsplitter was mounted on a post by epoxy to minimize strain and polarization distortion from strain. Samples were mounted in a Montana cryostation (Standard series). An out-of-plane magnetic field was applied through a home-made coil surrounding the sample chamber with a maximum field strength of 20 Oe. The incident intensity of the probe beam on the sample was 0.09 μW/μm$^2$ for all the measurements presented in the text. For control experiment with different probe beam intensities, see supplementary note 9. A 2D electron-multiplying CCD camera (Princeton Instruments, ProEM 512x512) was used to detect reflected light for real-time imaging.

Measurement of magnetic properties away from $T_c$: The hysteresis loop of CrBr$_3$ monolayer (Fig. 1e) and the DC magnetic susceptibility of both the CrBr$_3$ monolayer and the thin bulk CrBr$_3$ (Fig. 1f) were obtained from the magnetization-magnetic field (m-h) dependence under an external magnetic field provided by the home-made coil. The magnetization of each flake was obtained by averaging the magneto-optical contrast over a 3×3 μm region in the center of the flake. For temperatures close to $T_c$, the m-h curve of monolayer CrBr$_3$ becomes ill-defined due to the stochastic critical spin fluctuations at macroscopic scale.

Extraction of the correlation functions from real-time images: The temporal and spatial correlation functions (Fig. 2 and 3) were obtained directly from the real-time images of the magnetization fluctuations following the definition given in the text. The average was calculated as a time average over 5000 frames (taken at 70 frames per second) at each temperature. We note that, owing to the importance of the boundary effect, common analysis methods that require spatial averaging, such as the Fourier transform[36], become not applicable. The time average we adopted in this study is closer to the original definition of the ensemble average given the uniformity in time at equilibrium, and can directly provide information on the effects of spatial inhomogeneity.

Unusually strong critical slowing-down in 2D CrBr$_3$: Although critical fluctuations in principle can occur at any timescale, experimentally observed critical dynamics in solid-state spin system is usually faster than a millisecond[37,38]. Our observation of critical fluctuations at macroscopic length- and time-scales is hence surprising. To further confirm the unusually strong critical slowing-down in 2D CrBr$_3$, we performed AC susceptibility measurement as an independent probe. AC susceptibility is commonly used to determine the relaxation time of a system and its critical dynamics[38,39]. Indeed, prominent critical slowing-down is observed in monolayer CrBr$_3$ with timescale over a hundred milliseconds, but not in a ~ 10 nm thin bulk reference sample (see supplementary note 8). The distinctive differences between the monolayer and bulk CrBr$_3$ samples clearly show the enhanced critical fluctuations in 2D. The consistency between AC susceptibility and real-time imaging confirms that the observed fluctuations are from the intrinsic critical behaviors of 2D CrBr$_3$.

Repeated and reversible switching of 2D CrBr$_3$: Measurement was performed to determine the state of a magnetic bit following the application of a square gate pulse that temporarily enables critical fluctuations. Gate pulses of variable amplitude and width were generated by a digital delay/pulse generator (Stanford Research DG535), which was triggered by a data acquisition (DAC) card (National instruments USB-6212). In Fig. 4d, the gate pulses have an amplitude 0.4 V and a pulse width 50 ms. The DAC card also triggered the electron-multiplying CCD camera to synchronize the electrical control and the MCD imaging. Measurement of the magnetization of the bit was made 200 ms after the start of each pulse to ensure that $V_g$ and hence critical fluctuations are turned off. The above process was repeated with periodicity of 300 ms to obtain measurement sequences as exemplified in Fig. 4d.

Supplementary Information for

Imaging and control of critical spin fluctuations in two-dimensional magnets

Chenhao Jin[*], Zui Tao, Kaifei Kang, Kenji Watanabe, Takashi Taniguchi, Kin Fai Mak[*], Jie Shan[*]

* Correspondence to: jinchenhao@cornell.edu, km627@cornell.edu, jie.shan@cornell.edu

1. Determination of the $CrBr_3$ layer number

2. Relation between optical contrast and the MCD signal

3. Optimizing the MCD signal through the hBN thickness

4. Equal-time spatial correlation function in monolayer $CrBr_3$

5. Monte Carlo simulation of Ising model with boundary

6. Optical switching of magnetic state in sample S1

7. Gate-tunable monolayer $CrBr_3$ device S2

8. Critical dynamics probed by AC susceptibility measurement

9. Evaluation of the effects of the probe light

10. Supplementary movies

# 1. Determination of the CrBr₃ layer number

Atomically thin CrBr₃ samples employed in this study are encapsulated with hexagonal boron nitride (hBN) on both sides to avoid sample degradation. The relatively thick top hBN layer makes it difficult to accurately measure the thickness of CrBr₃ samples by atomic force microscopy (AFM). We therefore determine their thickness from the contrast in the optical microscopy image before hBN encapsulation. Figure S1a shows the optical microscopy image of sample S1 investigated in the main text. For this purpose, we first calibrate the relation between optical contrast and layer number by making reference CrBr₃ flakes encapsulated by few-layer hBN and measuring their thickness using AFM outside the glovebox. Figure S1b shows the optical microscopy image of representative reference flakes with different thickness. The substrate and the configuration for taking optical microscopy images are identical to that for S1. The thickness and optical contrast in the green channel of all reference flakes are summarized in Fig. S1c, including two monolayers, three bilayers and one trilayer. We find a good linear dependence between the optical contrast and layer number, as expected for very thin flakes. The blue dashed line in Fig. S1c represents optical contrast of sample S1 in the green channel, from which we determine it to be a monolayer.

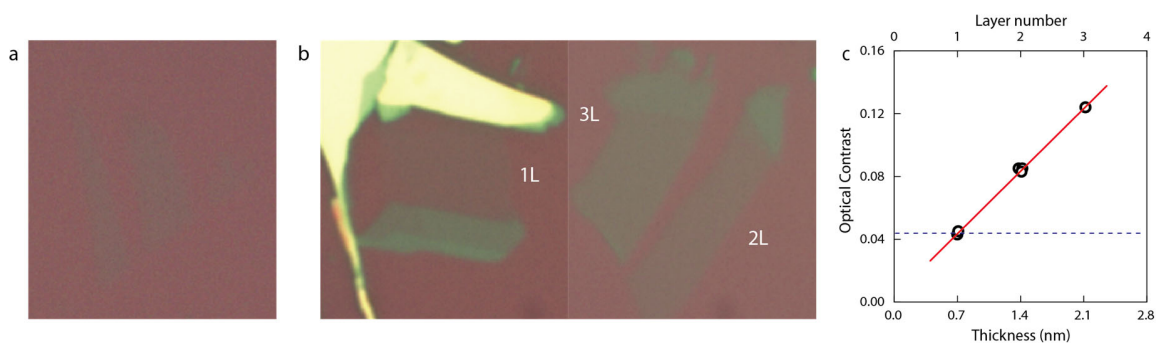

**Supplementary Figure S1.** (**a**) Optical microscopy image of sample S1 studied in the main text before hBN encapsulation. (**b**) Optical microscopy image of representative reference samples taken under the same experimental conditions as for sample S1. (**c**) Optical contrast in the green channel and AFM-determined thickness of reference samples show a good linear dependence. Blue dashed line represents the optical contrast of sample S1.

# 2. Relation between optical contrast and the MCD signal

To resolve a small magnetic circular dichroism (MCD) signal, we use linearly-polarized light (without loss of generality, along the $x$ direction) as the illumination source and analyse its polarization change induced by the sample. If the sample has zero magnetization $M$, the reflected light under normal incidence should have the same linear polarization as the incident light, and can be decomposed into left-circularly-polarized (LCP) and right-circularly-polarized (RCP) light with equal amplitude and phase:

$$\boldsymbol{E}_r(M=0) = E_{r0}\begin{pmatrix}1\\0\end{pmatrix} = \frac{E_{r0}}{2}\begin{pmatrix}1\\i\end{pmatrix} + \frac{E_{r0}}{2}\begin{pmatrix}1\\-i\end{pmatrix} \quad (1)$$

Here $E_{r0}$ is the amplitude of the reflected light field; the two components in the bracket represent light field along the $x$ and $y$ directions. If the sample has finite magnetization, the MCD and Kerr rotation are determined by an amplitude and phase difference between the LCP and RCP components, respectively. The reflected light field therefore becomes:

$$\boldsymbol{E}_r(M) = \frac{E_{r0}}{2}(1-\alpha)\begin{pmatrix}1\\i\end{pmatrix}e^{-i\theta} + \frac{E_{r0}}{2}(1+\alpha)\begin{pmatrix}1\\-i\end{pmatrix}e^{i\theta} \approx E_{r0}\begin{pmatrix}1\\\theta-i\alpha\end{pmatrix} \quad (2)$$

where $\alpha$ and $\theta$ are the MCD and Kerr rotation angles, respectively. We can see that both the MCD and Kerr rotation will induce a small field along the $y$ direction, perpendicular to the original linear polarization. To enhance the small perpendicular component, we interfere it with the parallel component. This can be done by placing the second polarizer (i.e. the analyser) at angle $\phi$ to the $y$ direction. After this polarizer, the final amplitude of the light field will be:

$$E_f/E_{r0} = \sin\phi + e^{i\delta}(\theta - i\alpha)\cos\phi \approx \phi + e^{i\delta}(\theta - i\alpha) \quad (3)$$

Here $e^{i\delta}$ is an additional phase difference between the $x$ and $y$ components of light that can be tuned by adding a waveplate before the analyser. For example, $\delta = 0$ without any waveplate; $\delta = \pi/2$ with a quarter-wave plate (QWP). By choosing a proper $\delta$ one can extract either the MCD or the Kerr rotation signal selectively. In our experimental configuration we choose $\delta = \pi/2$ to focus on the MCD signal (see Fig. 1a of the main text). The final light intensity transmitting the analyser will be:

$$I_f/I_{r0} = (\phi + \alpha)^2 + \theta^2 + X \quad (4)$$

Here $X$ is the inverse of the extinction ratio, representing the polarization-uncontrolled part of light. When the MCD and Kerr rotation are small, we can keep only the first-order term. The optical contrast induced by the magnetization $M$ of the sample is therefore:

$$\frac{I_f(M) - I_f(M=0)}{I_f(M=0)} \approx \frac{(\phi^2 + 2\phi\alpha + X) - (\phi^2 + X)}{\phi^2 + X} = \frac{2\phi}{\phi^2 + X}\alpha \quad (5)$$

We find that the optical contrast is linearly proportional to the MCD signal $\alpha$. Furthermore, by choosing $\phi \sim \sqrt{X}$, one can enhance the MCD signal by $\sim 1/\sqrt{X}$ times. In the experimental configuration, $\phi$ is controlled conveniently by a half-wave plate (HWP in Fig. 1a) in combination with a fixed analyser instead of controlling the analyser directly. We use $\phi = 0.6°$ for all measurements, which enhances the MCD signal by more than 100 times.

We note that, the magnetization $M$ of the sample directly affects its dielectric function (or equivalently, the optical conductivity). However, for a thin sample embedded in a multi-layer structure, the MCD signal will depend not only on its dielectric function, but also the local field factor of the environment. The next section will discuss the determination of the MCD signal $\alpha$ from both the dielectric function of the sample and the local field factor.

# 3. Optimizing the MCD signal through the hBN thickness

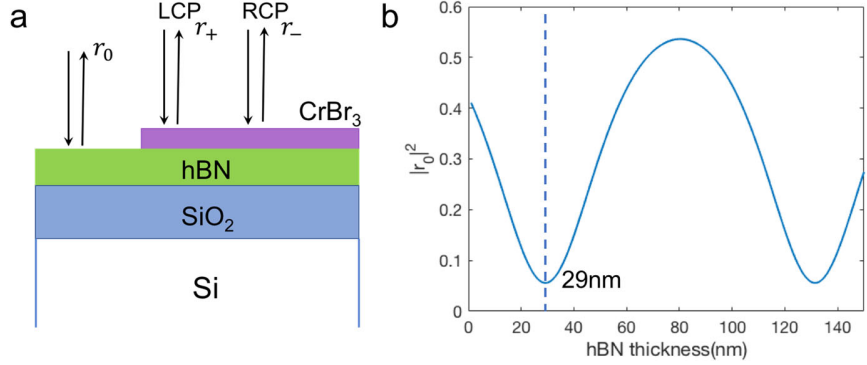

**Supplementary Figure S2.** (**a**) Multilayer structure under consideration. Upward and downward arrows represent incident and reflected light, respectively. (**b**) Simulated reflectance $|r_0|^2$ of hBN-SiO$_2$-Si multilayer with a 300-nm SiO$_2$ as a function of hBN thickness for 409-nm light.

For simplicity, we consider the structure in which CrBr$_3$ sits on top of a hBN-SiO$_2$-Si multilayer, as shown in Fig. S2a. The total reflection is the sum of multiple reflections at different interfaces. In the 2D limit, the contribution from the CrBr$_3$ layer to the total reflection is weak and can be treated as a perturbation. Using the transfer-matrix method for multilayer structures, we obtain the total reflection coefficient $r_+$ for LCP light and $r_-$ for RCP light:

$$r_\pm = r_0 + \Delta r_\pm \approx r_0 - \frac{i\pi d}{\lambda}[(1-r_0)^2 - C_l^2 \epsilon_\pm] \quad (6)$$

where $r_0$ is the reflection coefficient of the bare hBN-SiO$_2$-Si structure without CrBr$_3$, $\Delta r_\pm$ the contribution of CrBr$_3$, $C_l = (1 + r_0)$ the local field factor, $\lambda$ the wavelength of light in vacuum, and $d$ the thickness of CrBr$_3$. With finite magnetization, the CrBr$_3$ layer can respond differently to LCP and RCP light, as described by the different dielectric function $\epsilon_\pm$, respectively. To obtain $\epsilon_\pm$, we start with the general dielectric tensor for a 2D system with three-fold rotational symmetry and magnetization parallel to the z-axis (out-of-plane):

$$\epsilon = \begin{pmatrix} \epsilon_{xx} & \epsilon_{xy} & 0 \\ -\epsilon_{xy} & \epsilon_{xx} & 0 \\ 0 & 0 & \epsilon_{zz} \end{pmatrix}$$

We find $\epsilon_\pm = \epsilon_{xx} \pm i\epsilon_{xy}$. Due to the difference between $\epsilon_+$ and $\epsilon_-$, LCP and RCP light reflection can have different amplitude and/or phase.

The MCD signal $\alpha$ characterizes the relative amplitude difference (equivalent to the intensity difference by a factor of 2) between the LCP and RCP reflection,

$$\alpha = \frac{|r_+|-|r_-|}{|r_+|+|r_-|} = \frac{1}{2}\frac{I_+ - I_-}{I_+ + I_-}$$

Expanding $\alpha$ to the leading order in $(\Delta r_+ - \Delta r_-)$, we have

$$\alpha \approx \frac{\text{Re}[r_0^*(\Delta r_+ - \Delta r_-)]}{2|r_0|^2} = \text{Re}[\frac{\Delta r_+ - \Delta r_-}{2r_0}]$$

$$= \frac{\pi d}{\lambda}\text{Re}[i\frac{C_l^2}{2r_0}(\epsilon_+ - \epsilon_-)] = -\frac{\pi d}{\lambda}\text{Re}[\frac{C_l^2}{r_0}\epsilon_{xy}] \quad (7)$$

Similarly, Kerr rotation corresponds to the phase difference between the LCP and RCP reflection,

$$\theta = [\text{Arg}(r_+) - \text{Arg}(r_-)]/2$$

and is given by

$$\theta = \text{Im}\left[\frac{\Delta r_+ - \Delta r_-}{2r_0}\right] = -\frac{\pi d}{\lambda}\text{Im}[\frac{C_l^2}{r_0}\epsilon_{xy}] \quad (8)$$

The above expressions for $\alpha$ and $\theta$ suggest that a smaller $|r_0|$ generally leads to a larger MCD and Kerr signal. This can be intuitively understood since the total reflection contains components from both the substrate and the CrBr$_3$ sample, and only the latter contributes to the MCD and Kerr signal. As a result, one can enhance the measured MCD signal by choosing an optimal hBN thickness to minimize $|r_0|$ for the given light wavelength.

Figure S2b shows the dependence of $|r_0|^2$ on hBN thickness for a 300-nm SiO$_2$/Si substrate at wavelength of 409 nm. From the plot, we can see that the first minimum occurs at around 29 nm. In practice, the CrBr$_3$ sample is encapsulated with hBN on both sides. Following the above argument, the total thickness of the top and bottom hBN layers should be around 29 nm to minimize the reflection from the bare hBN-SiO$_2$-Si structure. For the monolayer CrBr$_3$ flake discussed in the main text, we measure the total hBN thickness to be 30 nm, which is close to the optimal value and enhances the MCD signal by several times compared to the case without hBN.

### 4. Equal-time spatial correlation function in monolayer CrBr$_3$

In the main text we have shown the equal-time spatial correlation function for monolayer CrBr$_3$ at several representative temperatures. In Fig. S3a and S3b, respectively, we show the complete temperature set of the spatial correlation function for point P1 and P2 as defined in the main text. We note that, for temperature around 21.8 K, the fluctuation dynamics is very slow (see Fig. 2e in the text). As a result, the obtained spatial correlation function is not as reliable due to insufficient averaging and non-ergodicity in the timescale of our measurement. On the other hand, the different spatial patterns between the two points can be clearly observed below T$_c$ for temperatures higher than 21.9 K.

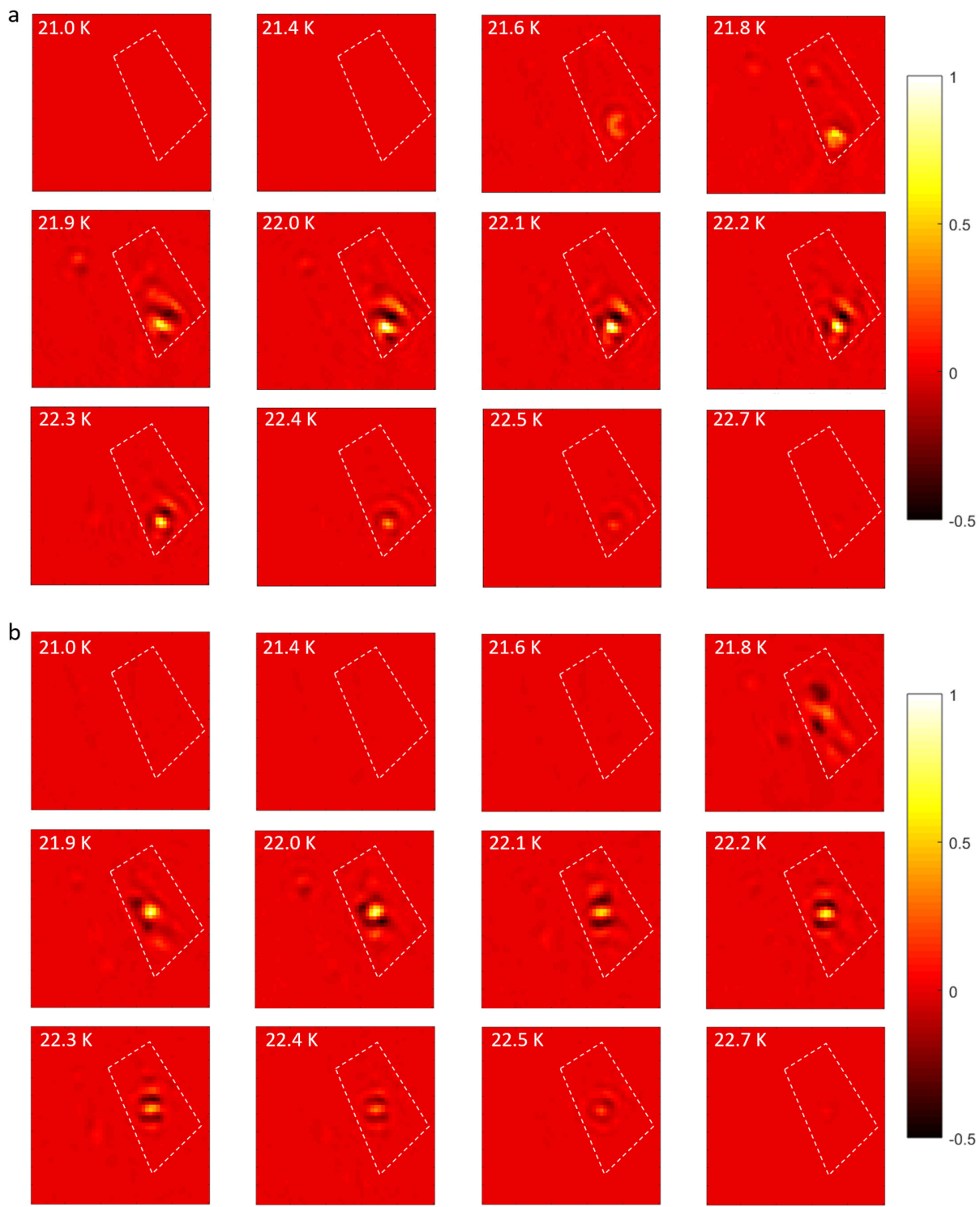

**Supplementary Figure S3.** Temperature-dependent spatial correlation functions for point P1 (**a**) and P2 (**b**) of the monolayer $CrBr_3$ flake studied in the main text.

## 5. Monte Carlo simulation of Ising model with boundary

We investigate the boundary effect on critical fluctuations in the 2D Ising model of a finite-size system (Fig. S4) by Monte Carlo simulation. The parallelogram corresponds to a triangular spin lattice with 130×50 sites. The orange and black regions represent sites inside and outside the sample, respectively. We adopt the boundary condition that no spin exists outside the boundary, and spins at the boundary will have a correspondingly reduced number of neighbouring sites. The configuration in Fig. S4 simulates a flake with finite size and quadrilateral shape; and the four interior angles of the quadrilateral are 60, 120, 90 and 90 degrees.

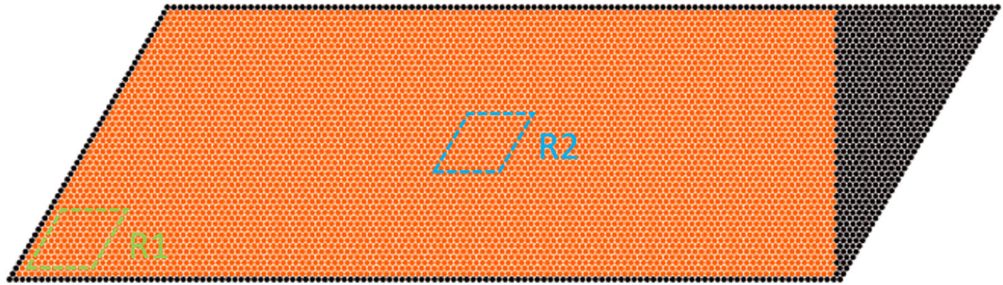

**Supplementary Figure S4.** Finite-size spin lattice for Monte Carlo simulation. Orange and black regions are inside and outside the sample. Green and blue dashed lines encircle two regions at a corner and in the center of the sample, respectively.

The boundary effect can be probed through comparing the temperature-dependent behaviours between regions at the 60-degree corner (R1, green dashed line in Fig. S4) and in the center (R2, blue dashed line), both containing 11×11 sites. At each temperature, the system is initialized to be spin up on all sites and evolves for $5\times10^6$ iterations. Afterwards, the system is considered to have reached equilibrium and the statistics (the average magnetization, fluctuation amplitude, temporal- and spatial-correlation function, etc.) is evaluated from the time average over the next $5\times10^6$ iterations. The data analysis is done in exactly the same way as in the experiment. The number of iterations is chosen due to the practical limits in computation time and storage space.

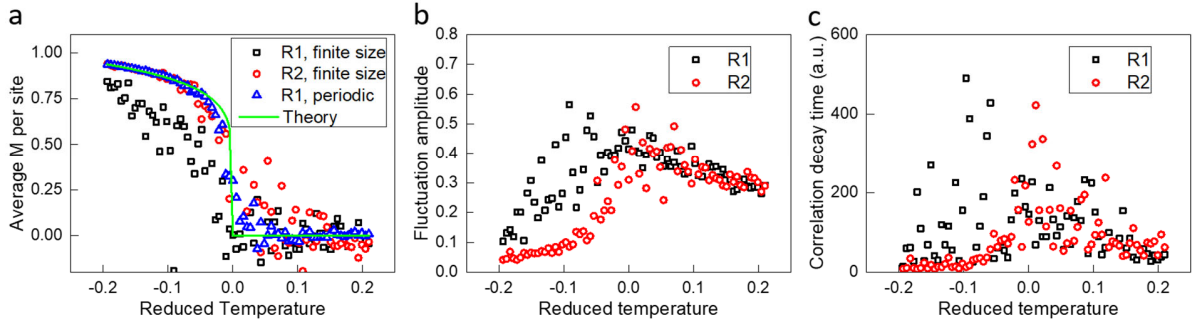

**Supplementary Figure S5.** Comparison between regions at the corner (R1, black symbols) and in the center (R2, red symbols). (**a**)-(**c**) Temperature-dependent average magnetization (**a**), spin fluctuation amplitude (**b**), and correlation decay time (**c**). Region R2 shows the large-scale fluctuations and the slowing-down behaviour that are peaked at $T_c$; while both behaviours are shifted towards lower temperatures for region R1. The reduced temperature is defined as $t = (T - T_c)/T_c$.

Figure S5a shows the average magnetization per site for region R1 (black) and R2 (red) at different temperatures. As a reference, we also provide results from region R1 in a similar spin lattice but without boundary (blue, obtained from the 130×50 lattice shown in Fig. S4 with periodic boundary conditions) and the theoretically predicted power-law behaviour (green). The reduced temperature here is defined as $t = (T - T_c)/T_c$. Both region R1 with periodic boundary conditions (blue) and region R2 in the finite size case (red) match reasonably well with the theoretical prediction. In contrast, region R1 in the finite size case (black) shows a substantially smaller average magnetization in a large temperature range below $T_c$. This can be attributed to the boundary-facilitated large-scale fluctuations: As shown in Fig. S5b, the fluctuation amplitude of region R2 is peaked at $T_c$; while for region R1 large fluctuations emerge at lower temperatures. Such behaviour is consistent with our experimental observation and can be intuitively understood from the smaller number of neighbouring sites for regions at the boundary.

Furthermore, we can calculate the temporal correlation function $C(\mathbf{r}, \Delta t) = \langle M(\mathbf{r}, t) M(\mathbf{r}, t + \Delta t) \rangle - \langle M(\mathbf{r}, t) \rangle \langle M(\mathbf{r}, t + \Delta t) \rangle$ and extract the correlation decay time following the same way as in the experiment, see Fig. S5c. Indeed, slowing down of the fluctuation dynamics exhibits close relation to the increase of the fluctuation scale (Fig. S5b). As a result, region R1 and R2 show prominent slowing-down behaviours below and at $T_c$, respectively, supporting the important role of the boundary effect in determining the critical dynamics.

Similarly, we can obtain the spatial correlation function $C(\mathbf{r_1}, \mathbf{r_2}) = \langle M(\mathbf{r_1}, t) M(\mathbf{r_2}, t) \rangle - \langle M(\mathbf{r_1}, t) \rangle \langle M(\mathbf{r_2}, t) \rangle$ at representative temperatures, as shown in Fig. S6. Here $\mathbf{r_1}$ is mapped over the entire lattice and $\mathbf{r_2}$ is fixed at the center site of region R1 (Fig. S6a) and R2 (Fig. S6b), respectively. The number on the left of each panel labels the reduced temperature. Again strong spatial correlation of fluctuations is only observed near $T_c$ for region R2, similar to the critical behavior of an infinite and homogeneous system. Nevertheless, for region R1 large spatial correlation emerges and peaks below $T_c$, owing to the boundary effect.

We note that, the simple simulation here is to provide an intuitive and qualitative understanding of the boundary effect on the critical fluctuations in 2D magnets, and cannot be quantitatively compared to the experimental results without fully considering finite magnetic anisotropy, dipole-dipole interaction and other types of spatial inhomogeneity. For instance, part of the flake does not fluctuate at all over the entire temperature range (see Fig. 2a in the main text), which can constitute effective boundary. Therefore, in practice all regions of the flake may experience some extent of the boundary effect.

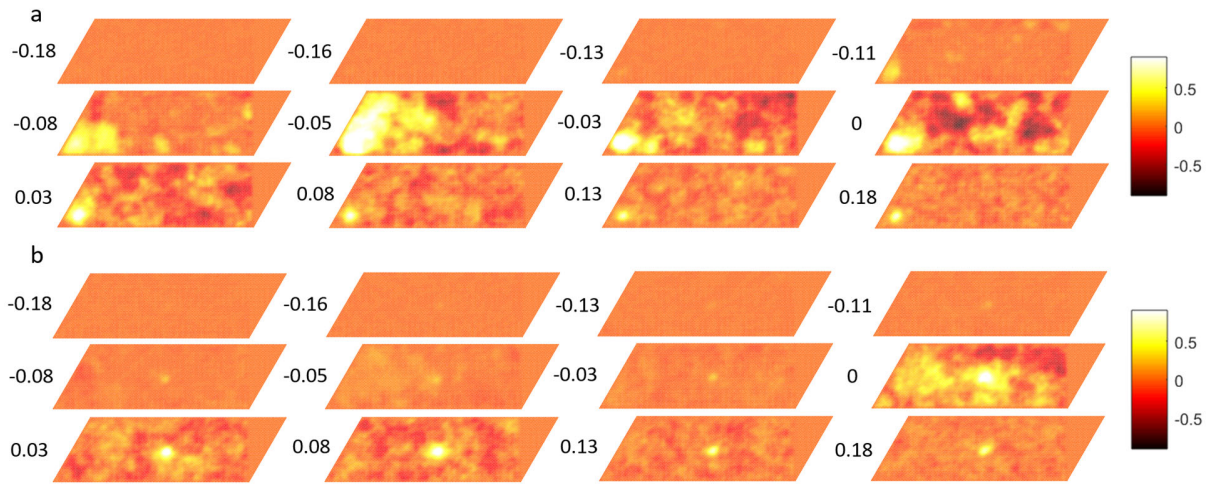

**Supplementary Figure S6.** Spatial correlation function $C(\boldsymbol{r}_1, \boldsymbol{r}_2)$ at representative temperatures, where $\boldsymbol{r}_2$ is fixed at the center site of region R1 (**a**) and R2 (**b**), respectively. The numbers on the left of each image denote the reduced temperature.

## 6. Optical switching of magnetic state in sample S1

Because the critical fluctuations are extreme sensitivity to the environment, we can use light to switch the magnetic state in sample S1, similar to the electrical control of critical fluctuations demonstrated in device S2 (Fig. 4 in the text). We use light at 532 nm to control the fluctuations in sample S1 at a base temperature of 21.6 K through heating and/or photo-doping effects. The 532nm control light is combined with the probe beam by another beamsplitter (not shown in Fig. 1a) before polarizer 1. Afterwards, the control beam followed the same beam path as the probe beam and was filtered out by a band pass filter after polarizer 2. The incident intensity of the control beam on the sample was 0.9 µW/µm$^2$.

Figure S7a shows the light pulse sequence as well as the magnetization evolution (black) of position P4 on the flake (labelled by arrows in Fig. S7b). When control light is off (unshaded regions in Fig. S7a), the critical fluctuations are absent, and the magnetization stays either positive (state 1) or negative (state 0), as imaged in the left and right panels of Fig. S7b. When control light is on (yellow shaded regions in Fig. S7a), the fluctuations are activated. Using real-time feedback control of light (i.e. removing it at the right moment according to the real-time magnetization measurement), the system can be driven into the desired state(s) in a non-volatile way (see supplementary movie 14).

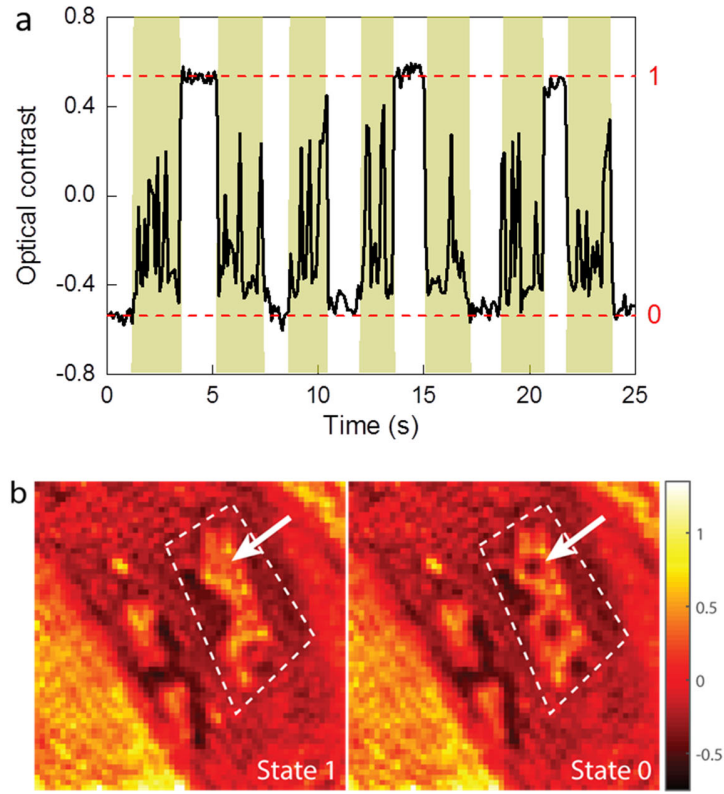

**Supplementary Figure S7.** Switching magnetic states by optically controlling the critical fluctuations. **a**, Magnetization time trace of point P4 (arrows in **b**). When control light is off (unshaded regions), the critical fluctuations are absent and the magnetization stays in either of the two stable states, state 1 and 0 (red dashed lines). When control light is on (yellow shaded regions), the critical fluctuations are present. Specific state can be written into the magnet by removing control light at the right moment according to the real-time magnetization measurements. **b**, MCD image of state 1 (left) and state 0 (right) for point P4 indicated by white arrows.

## 7. Gate-tunable monolayer CrBr$_3$ device S2

Figure S8a shows the schematic sideview of monolayer CrBr$_3$ device S2. Two bilayer graphene flakes are used as contact and back gate electrodes, respectively, and the bottom hBN of ~20nm thickness is used as the gate dielectric. A gate voltage $V_g$ can be applied across the capacitor formed by the two graphene flakes to control $T_c$ of monolayer CrBr$_3$. The whole stack is placed in the center of a metallic ring structure (Fig. S8, b and c), which can apply a local magnetic field for AC susceptibility measurement. Fig. S8d shows the AC susceptibility of the device measured with different gate voltages (see section 8 for measurement details). The critical temperature can be extracted from the maximum of the AC susceptibility (dashed line), from which we find a monotonic shift in $T_c$ of ~ -0.4 K/V. As a result, at a fixed sample temperature below $T_c$, the system's distance to the critical point can be either decreased or increased with positive or negative $V_g$, thereby enhancing or reducing critical fluctuations.

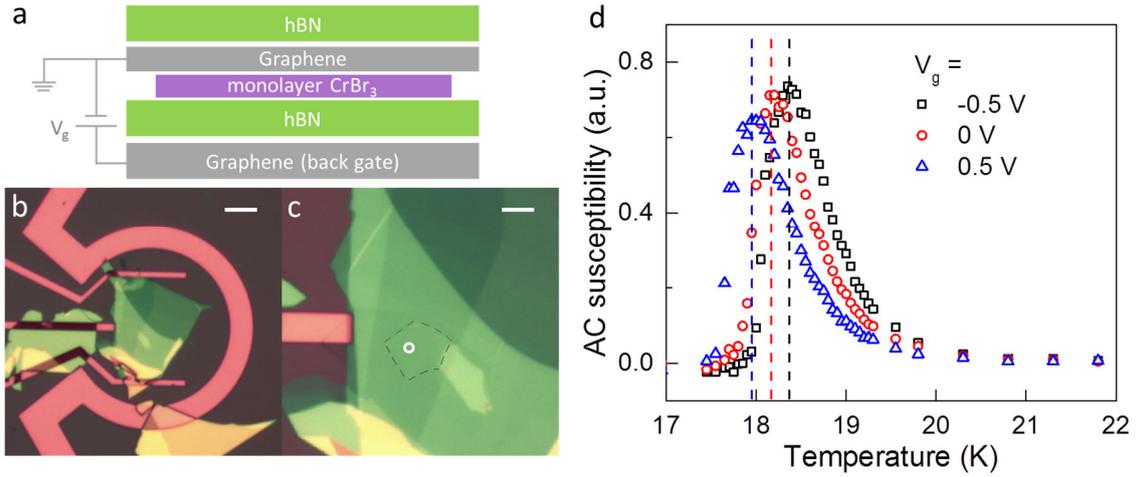

**Supplementary Figure S8.** Monolayer $CrBr_3$ device S2. **a**, Schematic side view of monolayer $CrBr_3$ device S2 with graphene contact and gate electrodes. The gate voltage $V_g$ is used to control critical fluctuations. **b,c**, Optical microscopy image of device S2 on metallic ring structure (**b**) and zoom-in at $CrBr_3$ flake (**c**). The gray dashed line in **c** labels contours of the $CrBr_3$ monolayer. Scale bars for (**b**) and (**c**) are 20 μm and 5 μm, respectively. **d**, Real part of AC susceptibility measured on position P3 (white circle in **c**) with gate voltage of -0.5 (black), 0 (red) and 0.5 (blue) volts. The gate voltage induces a monotonic shift in $T_c$ of ~ -0.4 K/V.

## 8. Critical dynamics probed by AC susceptibility measurement

To further confirm that the observed dynamics is from critical fluctuations, we perform magnetic AC susceptibility measurement as an independent probe. AC susceptibility is a commonly used technique to determine the relaxation time of a system and to probe its critical dynamics[37,38]. In this method, an oscillating magnetic field of frequency $\omega$ is applied and the induced magnetization is recorded with both amplitude and phase information. The in-phase and out-of-phase signal correspond to the real and imaginary part of the AC susceptibility ($\chi_1$ and $\chi_2$), respectively. If the system's dynamics is much faster than the period of the driving field, the response is instantaneous and $\chi_2$ vanishes. On the other hand, if the relaxation time of the system becomes comparable to the modulation period, $\chi_2$ becomes comparable to $\chi_1$. Quantitatively, the relaxation time can be evaluated as $\sim \frac{\chi_2}{\chi_1}\frac{1}{\omega}$ (Ref. [37,38]).

Figure S9a is the temperature dependent AC susceptibility measured on position P3 in sample S2. Here an oscillating magnetic field with an amplitude of 0.3 oe and a modulation frequency of 36 Hz is applied. To detect the in-phase component of the induced magnetization, we synchronize the CCD camera with the modulation field. To detect the out-of-phase component of the induced magnetization, we synchronize the CCD camera with a delayed modulation field by quarter a cycle. We observe a sharp increase of $\chi_1$ (black) when approaching $T_c$ (~ 18 K). In addition, we also observe an even sharper temperature dependence of $\chi_2$ (red), which exists only in a narrow temperature range of ~ 0.5 K around $T_c$. Note that $T_c$ of monolayer $CrBr_3$ varies from sample to sample since it depends sensitively on carrier density, strain, etc.

The relaxation time is evaluated in Fig. S9b. The system shows a critical slowing-down behavior in a narrow temperature range (~ 0.5 K) around $T_c$ with relaxation time over a hundred milliseconds. We note that the critical point for ferromagnetic phase transitions exists strictly only under zero external magnetic field. The presence of a small field for the AC susceptibility measurement increases the distance of the system to the critical point and partially suppresses the critical behavior. The measured relaxation time is therefore a lower bound. The result of Fig. S9b agrees well with the fluctuation dynamics directly obtained from the real-time imaging.

The observed slowing-down behavior in monolayer $CrBr_3$ in the close vicinity of $T_c$ is unique to 2D systems. In Fig. S9c we show the AC susceptibility of a thin bulk $CrBr_3$ sample (~10 nm thick) under the same experimental configuration. In sharp contrast to the case of monolayer $CrBr_3$, $\chi_2$ here is negligible in a large temperature range around $T_c$ (~ 30 K) and no critical slowing-down behavior is discernible within the experimental uncertainties (Fig. S9d). This result is also consistent with the absence of critical fluctuations in real-time imaging of bulk samples (supplementary movie 1).

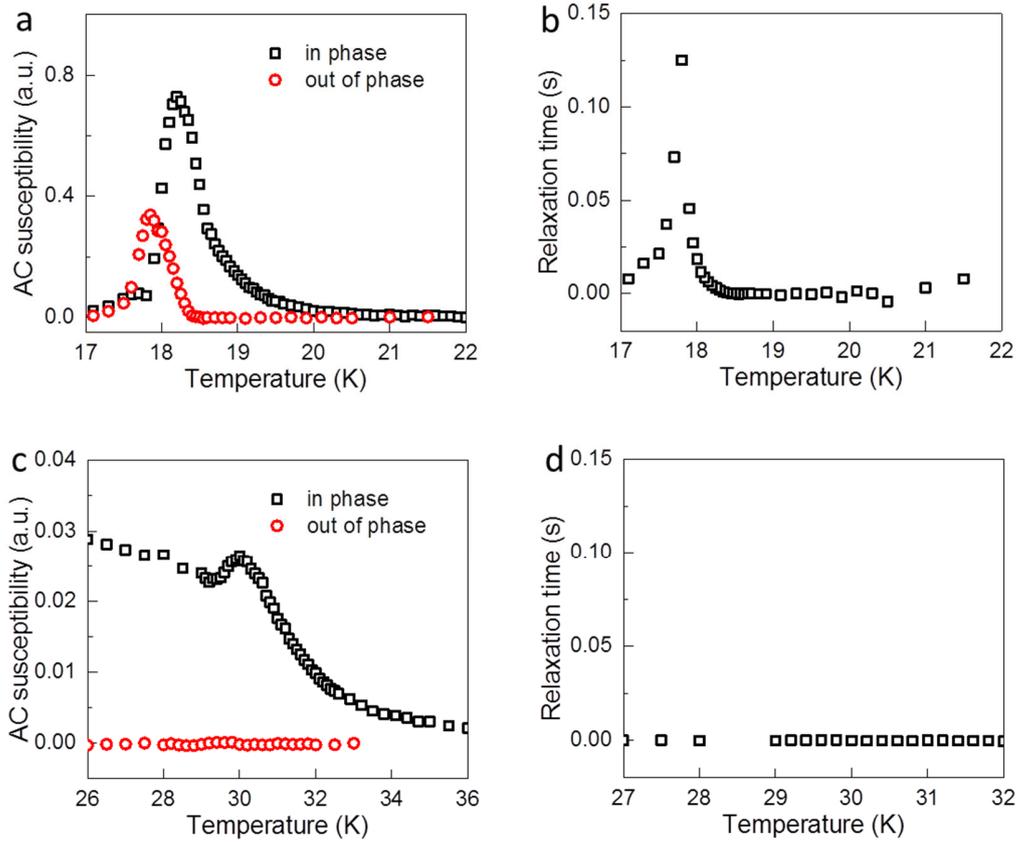

**Supplementary Figure S9.** Critical slowing-down probed by AC susceptibility measurement. **a**, AC susceptibility measured on monolayer CrBr$_3$ (position P3 on device S2) using an oscillating magnetic field with an amplitude of 0.3 oe and a modulation frequency of 36 Hz. The emergence of an out-of-phase component in the induced magnetization (red) in a narrow temperature range around $T_c$ (~18K) indicates that the relaxation time of the system becomes comparable to the modulation period. **b**, Relaxation time determined by comparing the in-phase and out-of-phase components of the induced magnetization. A prominent critical slowing-down behavior is observed, with relaxation time over a hundred milliseconds. This is consistent with the fluctuation dynamics obtained from real-time imaging. **c**, **d**, AC susceptibility measured on a thin bulk CrBr$_3$ sample (~10 nm thick) (**c**) and the extracted relaxation time (**d**). No discernible out-of-phase signal or critical slowing-down is observed within experimental uncertainties.

## 9. Evaluation of the effects of the probe light

To determine effects of the probe beam on critical fluctuations, we have performed the following control experiments:

(i)    We have studied the fluctuation behavior in monolayer CrBr$_3$ using different probe

intensities in real-time imaging;
(ii) We have studied the fluctuation behavior in monolayer $CrBr_3$ by measuring the magnetic AC susceptibility;
(iii) We have compared the fluctuation behavior of $CrBr_3$ samples of different thicknesses (monolayer vs thin bulk).

Figure S10 shows the time traces of MCD optical contrast from a monolayer $CrBr_3$ on silicon substrate (device S2, position P3) at different temperatures (temperatures are measured with probe light off). Three different probe intensities are used: 0.03 (Fig. S10a), 0.09 (Fig. S10b), and 0.35 µW/µm$^2$ (Fig. S10c). The corresponding temperature dependences of the fluctuation amplitude and correlation decay time are summarized in Fig. S10d and e, respectively. We find that different probe intensities do not affect the behaviors of critical fluctuations, except shifting the apparent "$T_c$" to a lower value for a higher probe intensity. This can be understood as the probe light induces an average temperature increase from homogenous heating (presumably through the Si substrate absorption). The temperature increase is estimated to be ~ 0.05 K for 0.09 µW/µm$^2$, which was used throughout the experiment in the main text. The absence of a probe intensity dependence of the behaviors of critical fluctuations indicates that the magnetization fluctuations will remain similar even if the probe light intensity is zero (at most with an overall temperature shift), thereby excluding the probe light as a potential origin of the observed MCD optical contrast fluctuations and dynamics.

In the second control experiment, we measure the magnetic AC susceptibility of monolayer $CrBr_3$ as a function of temperature (Fig. S9). The probe intensity was 0.09 µW/µm$^2$. As described above in section 8, the AC susceptibility measurement is based on lock-in detection, which sets a narrow bandpass filter centered at the frequency of the oscillating field and can effectively reject random fluctuations from probe laser or temperature instabilities. The consistency between the AC susceptibility measurement and the real-time imaging therefore further confirms that the observed fluctuations are intrinsically from the critical behavior of the sample.

Finally, we compare the fluctuation behavior of monolayer and thin bulk (~ 10 nm thick) $CrBr_3$ (see Fig. 2c and supplementary movie 1). Critical fluctuations with slowing-down dynamics are observed only in monolayer samples. If the probe laser noise were responsible for the observed effects, it would affect both monolayer and thin bulk $CrBr_3$. The absence of fluctuations in the think bulk samples again excludes the probe as the origin of the observed fluctuations.

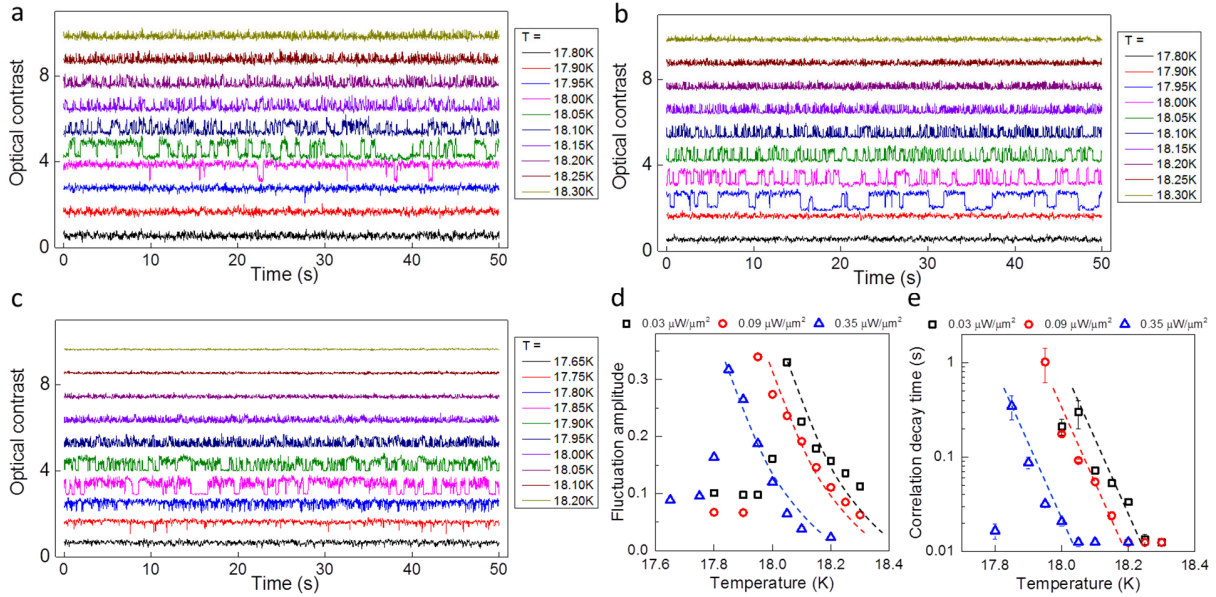

**Supplementary Figure S10.** Probe light intensity dependence. **a-c**, Magnetization time traces of a monolayer $CrBr_3$ device (position P3, device S2) measured with a probe light intensity of 0.03 (**a**), 0.09 (**b**) and 0.35 $\mu W/\mu m^2$ (**c**) at different temperatures. **d, e** Extracted fluctuation amplitude and correlation decay time from **a-c** as a function of temperature. The probe light leads to an overall temperature shift (~0.05 K for the experimentally used probe intensity of 0.09 $\mu W/\mu m^2$) from homogenous heating, and (after considering the temperature shift) does not change the fluctuation behavior, such as the amplitude or correlation decay time. Dashed lines are guide to the eye.

## 10. Supplementary movies

All supplementary movies were taken under zero total out-of-plane magnetic field, as determined by the center of the hysteresis curves. This was achieved by compensating the field from the environment with a home-made coil.

Supplementary movie 1: Real-time imaging of thin bulk $CrBr_3$ at temperature of 28 Kelvin. The movie was taken and shown both at 10 frames per second (fps).

Supplementary movie 2: Real-time imaging of monolayer $CrBr_3$ S1 at temperature of 21.0 Kelvin. The movie was taken at 70 fps and shown at 30 fps.

Supplementary movie 3: Real-time imaging of monolayer $CrBr_3$ S1 at temperature of 21.4 Kelvin. The movie was taken at 70 fps and shown at 30 fps.

Supplementary movie 4: Real-time imaging of monolayer $CrBr_3$ S1 at temperature of 21.6 Kelvin. The movie was taken at 70 fps and shown at 30 fps.

Supplementary movie 5: Real-time imaging of monolayer $CrBr_3$ S1 at temperature of 21.8 Kelvin. The movie was taken at 70 fps and shown at 30 fps.

Supplementary movie 6: Real-time imaging of monolayer $CrBr_3$ S1 at temperature of 21.9 Kelvin. The movie was taken at 70 fps and shown at 30 fps.

Supplementary movie 7: Real-time imaging of monolayer $CrBr_3$ S1 at temperature of 22.0 Kelvin. The movie was taken at 70 fps and shown at 30 fps.

Supplementary movie 8: Real-time imaging of monolayer $CrBr_3$ S1 at temperature of 22.1 Kelvin. The movie was taken at 70 fps and shown at 30 fps.

Supplementary movie 9: Real-time imaging of monolayer $CrBr_3$ S1 at temperature of 22.2 Kelvin. The movie was taken at 70 fps and shown at 30 fps.

Supplementary movie 10: Real-time imaging of monolayer $CrBr_3$ S1 at temperature of 22.3 Kelvin. The movie was taken at 70 fps and shown at 30 fps.

Supplementary movie 11: Real-time imaging of monolayer $CrBr_3$ S1 at temperature of 22.4 Kelvin. The movie was taken at 70 fps and shown at 30 fps.

Supplementary movie 12: Real-time imaging of monolayer $CrBr_3$ S1 at temperature of 22.5 Kelvin. The movie was taken at 70 fps and shown at 30 fps.

Supplementary movie 13: Real-time imaging of monolayer $CrBr_3$ S1 at temperature of 22.7 Kelvin. The movie was taken at 70 fps and shown at 30 fps.

Supplementary movie 14: Real-time imaging of magnetic switching in monolayer $CrBr_3$ S1 by controlling the critical fluctuations with light. The movie was taken and shown both at 10 fps.